%% file: ms.tex
\documentclass{article}
\usepackage{spconf,amsmath,graphicx}
\usepackage{amssymb}  % assumes amsmath package installed
\usepackage[utf8]{inputenc}
\usepackage[amsmath,thmmarks,hyperref]{ntheorem} % nicer theorems
\usepackage{bbm}               % nicer bbm fonts
\usepackage{mathtools}         % some nice tricks for stackrel
\usepackage{stmaryrd}          % more brackets
\usepackage{paralist}          % better enumerate lists
\usepackage{aliascnt}          % TEST alias counter for ntheorem environments
\usepackage{subcaption} %for subfigures
\usepackage{mathrsfs}
\usepackage{color}

\usepackage{textcomp}

\usepackage{booktabs}
\input{Macros}
\graphicspath{{pics/}}

\title{Dynamical Component Analysis (DyCA) and its Application on epileptic EEG} 
\name{Katharina Korn and Bastian Seifert and Christian Uhl
\thanks{The authors acknowledge funding by the European Regional Development
Fund (ERDF) and the support by the Bayerische Forschungsstiftung
within the project Nilpherd. We thank the Epilepsy Centre at the
Department of Neurology, Universitätsklinikum Erlangen for provided
data and BESA GmbH and Epilepsy Centre for fruitful ideas and
discussions.}}
\address{Ansbach University of Applied Sciences\\
  Faculty of Engineering Sciences\\
  Residenzstr.~8, 91522 Ansbach, Germany \\
  \{katharina.korn,bastian.seifert,christian.uhl\}@hs-ansbach.de}
\begin{document}

% \ninept
%
\maketitle

\begin{abstract}
    Dynamical Component Analysis (DyCA) is a recently-proposed method to detect
    projection vectors to reduce the dimensionality of multi-variate
    deterministic datasets. It is based on the solution of a
    generalized eigenvalue problem and therefore straight forward to
    implement. DyCA is introduced and applied to EEG data of epileptic
    seizures. The obtained eigenvectors are used to project the signal
    and the corresponding trajectories in phase space are compared
    with PCA and ICA-projections. The eigenvalues of DyCA are utilized
    for seizure detection and the obtained results in terms of
    specificity, false discovery rate and miss rate are compared to
    other seizure detection algorithms.
\end{abstract}
\section{Introduction}
\label{sec:introduction}%

Automatic detection of epileptic events in EEG data is a challenging
problem. On the one hand the detection of all epileptic events,
especially during live-monitoring sessions, is desireable. On the
other hand not too many epilepsy alarms should be triggered. This
represents a common classification problem aiming at high detection
rate as well as a high specificity.

Current approaches to solve this problem use very sophisticated
techniques, like the combination of wavelet transform with classical
machine learning classification approaches
\cite{Ahmadi:2017,Acharya:2012PCA,Acharya:2012ICA} or instead with a
deep convolutional neural network \cite{Acharya:2018}.

Dimensionality reduction of deterministic multi-variate time-series is
another ambitious problem. Most of the currently available tools of
dimensionality reduction, like PCA~\cite{Pearson:1901} or
ICA~\cite{Hyvaerinen.Oja:2000a} and modifications thereof, make a
\emph{stochasticity} assumption on the time-series on which they can
be applied. Presumably due to the lack of better matching techniques
they are often used for dimensionality reduction of deterministic
time-series even if its assumptions are not fulfilled.
Recently~\cite{Seifert.Korn.Hartmann.Uhl:2018a} the authors proposed a
new method for dimensionality reduction of deterministic time-series:
dynamical component analysis (DyCA). This method relys on a
determinacy assumption on the time-series. The projection onto a
lower-dimensional space is then found by solving a generalized
eigenvalue problem.

The eigenvalues of the generalized eigenproblem of DyCA measure
the quality of the assumption of linear determinism for the investigated
data. For certain conditions the corresponding eigenvectors together with
some linear transforms yield an optimal projection to represent the signal
by a deterministic non-linear differential equation.

Since EEG data during an epileptic event is known to be of
deterministic structure~\cite{vanVeen.Liley:2006}, while the EEG data
during normal activity is of stochastic nature, we assume that the
detection of epileptic events might be possible by investigating the
generalized eigenvalues of DyCA. In this article we examine this
approach and demonstrate the power of DyCA with respect to the
obtained trajectories in phase space by projecting the original signal
onto the DyCA eigenvectors. The DyCA eigenvalues are utilized to
implement a novel seizure detection algorithm and its results in terms
of specificity, false discovery rate and miss rate are compared to
other studies \cite{Acharya:2012PCA,Acharya:2012ICA,Acharya:2018}.

The paper is structured as follows. In Section~\ref{sec:DyCA} we present
the basic concepts underlying DyCA and demonstrate the formulation of
the dimensionality reduction process.
Section~\ref{sec:EpilepsyDeterminism} deals with the assumption and
confirmation of a deterministic model of EEG data of epileptic
seizures. In Section~\ref{sec:SeizureDetection} a DyCA-based method for
detection of epileptic seizures is evaluated and compared. Finally the
results are discussed (Section~\ref{sec:Discussion}) and concluded
(Section~\ref{sec:Conclusion}).

\section{Dynamical Component Analysis (DyCA)}
\label{sec:DyCA}%

Dynamical Component Analysis (DyCA) is a
recently-proposed~\cite{Seifert.Korn.Hartmann.Uhl:2018a} method for
dimensionality reduction of deterministic time-series and can be derived
as follows. Assume, given a high-dimensional deterministic time-series
$q(t) \in \mathbb{R}^N$ with dynamics governed by a
low-dimensional system of ordinary differential equations, the signal
can be decomposed into
\begin{equation}
    \label{eq:DecompositionHighDimSignal}
    q(t) = \sum_{i=1}^{n} x_i(t) w_i
\end{equation}
using time-dependent amplitude $x_i(t)$ and vectors
$w_i \in \mathbb{R}^N$ with $n \ll N$. The amplitudes are then assumed
to be governed by a set of ordinary differential equations, divided into
a set of linear differential equations,
\begin{equation}
    \label{eq:LinearEquationsDynamics}
    \begin{split}
        \dot{x}_1 &= \sum_{k=1}^n a_{1,k} x_k \\
        &\vdots \\
        \dot{x}_m &= \sum_{k=1}^n a_{m,k} x_k , 
    \end{split}
\end{equation}
and a set of non-linear differential equations with smooth functions~$f_i$: 
\begin{equation}
    \label{eq:NonlinearEquationsDynamics}
    \begin{split}
        \dot{x}_{m+1} &= f_{m+1}(x_1,\dots,x_n) \\
        &\vdots \\
        \dot{x}_n &= f_n(x_1,\dots,x_n).
    \end{split}
\end{equation}
Furthermore we assume that $m \geq n/2$, i.e.\, there exist more linear
than non-linear equations, and that every amplitude $x_i$ associated
to a non-linear equation appears in the right-hand side of at least one of
the linear equations without knowing the coefficients $a_{i,k}$ or the smooth
functions $f_i$.

Then projection vectors, $u_i, v_j \in \mathbb{R}^N $, can be found
containing the dynamics by minimizing the cost function
\begin{equation}
    \label{eq:DyCACostFunction}
    D(u,v,a) = \frac{\timeavg{\norm{\dot{q}^\top u - \sum_j a_j q^\top
          v_j}_2^2}}{\timeavg{\norm{\dot{q}^\top u}_2^2}},
\end{equation}
where $\timeavg{\argument}$ denotes the average over time. The rational behind
this is that at a minimum of $D$ all the information on how to project
onto the non-linear parts is contained in $\sum_j a_j q^\top v_j$ and
for the linear parts it is contained in $\dot{q}^\top u$.

The minima of $D$ for the vectors $u$ can be determined by a
generalized eigenvalue problem
\begin{equation}
    \label{eq:DyCAEigenproblem}
    C_1 C_0^{-1} C_1^\top u = \lambda C_2 u,
\end{equation}
with correlation matrices
$C_0 = \timeavg{q q^\top}, C_1 = \timeavg{\dot{q} q^\top}$, and
$C_2 = \timeavg{\dot{q} \dot{q}^\top}$. Furthermore there exists the
connection $u = \lambda C_2^{-1} C_1 \sum_j a_j v_j$. Thus by
projecting onto
\begin{equation}
    \label{eq:DyCAProjectionSpace}
    \mathsf{span}\{u_1, \dots, u_m, C_1^{-1} C_2 u_1, \dots, C_1^{-1}
    C_2 u_m\}
    = \mathbb{R}^n
\end{equation}
all relevant information is received.

On the other hand, the eigenvalues of the generalized
eigenproblem~\eqref{eq:DyCAEigenproblem} reveal something more, as then
the cost function~\eqref{eq:DyCACostFunction} takes the value
\begin{equation}
    \label{eq:MinValueDyCACostFunction}
    D_{\min} = 1 - \lambda. 
\end{equation}
Thus the number of the generalized eigenvalues with a value of approximately $1$
are a measure of the number of linear equations contained in the data. In the
subsequent sections this connection will be exploited to detect regions in
time-series with highly deterministic parts, like epileptic seizures. 

\section{Epilepsy - deterministic EEG data}
\label{sec:EpilepsyDeterminism}%

Unlike a first thought might suggest, during epileptic events the EEG
data is much more regular than during normal phases. Indeed there are
even models suggesting Shilnikov chaos to appear during epileptic
seizures~\cite{vanVeen.Liley:2006,Friedrich.Uhl:1996a}. In its easiest
form a system showing Shilnikov chaos can be described by a set of
three differential equations of the form
\begin{equation}
    \label{eq:ShilnikovChaos}
    \begin{split}        
        \dot{x}_1 &= x_2 \\
        \dot{x}_2 &= x_3 \\
        \dot{x}_3 &= f(x_1,x_2,x_3),
    \end{split}
\end{equation}
with a non-linear smooth function $f$. Thus one can assume that the
assumptions of DyCA are fulfilled and DyCA can be applied to epileptic
EEG data.

\begin{figure}
    \centering
    \includegraphics[width=0.45\textwidth]{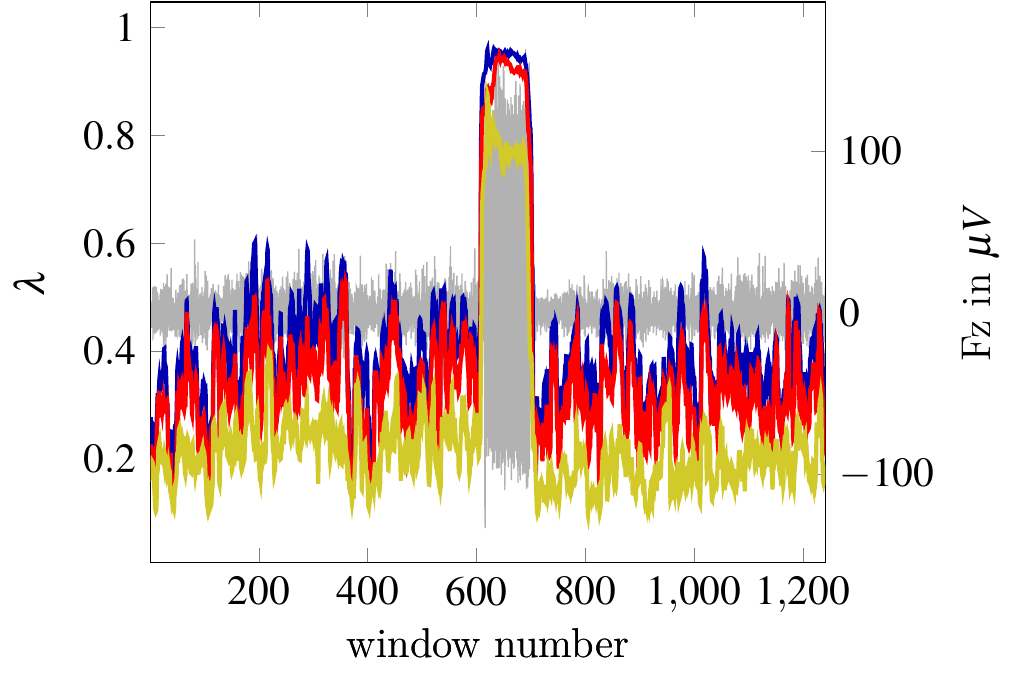}
    \caption{The three largest eigenvalues of DyCA on moving windows
      of EEG data. In the background the Fz-electrode is shown in
      grey (right axis). }
    \label{fig:WindowedDyCAEpilepsy}
\end{figure}

Fig.~\ref{fig:WindowedDyCAEpilepsy} presents the three largest eigenvalues
of DyCA applied to a moving window of an EEG dataset with an epilectic seizure.
Each investigated window has a length of three seconds and 90\,\% overlap. An epileptic seizure occurs
in-between window number 600 to approx. 700.

\begin{figure*}
    \centering
    \includegraphics[width=\textwidth]{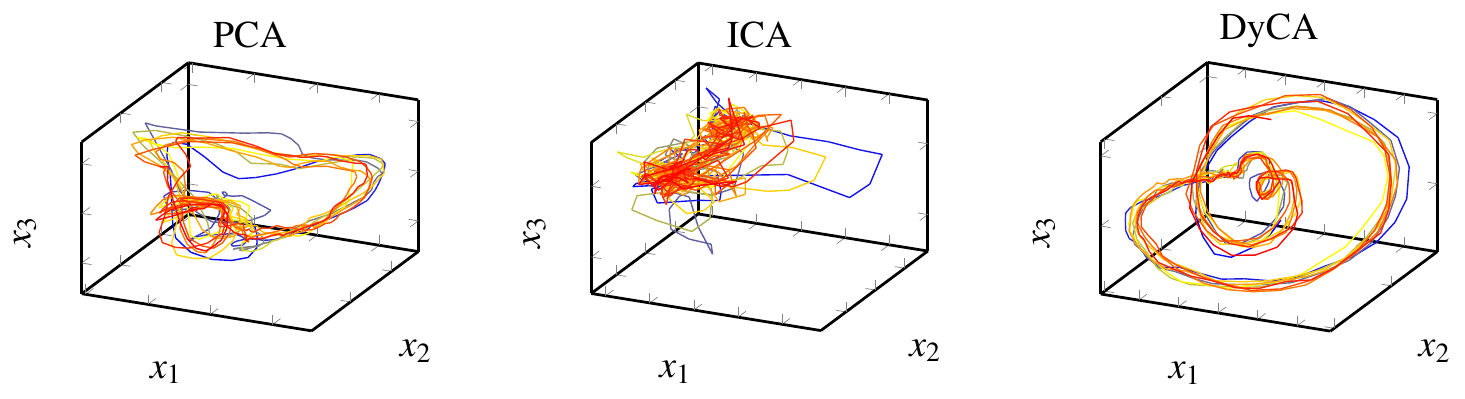}
    \caption{The projected trajectory of DyCA compared with PCA and
      ICA trajectories. Only the DyCA projection resembles a
      homoclinic orbit of Shilnikov chaos. The color indicates the
      time evolution.}
    \label{fig:ShilnikovChaosHomoclinic}
\end{figure*}

A clear jump of the first three eigenvalues is observed during seizure.
This confirms the assumed low-dimensional deterministic behaviour in ictal
phases of the signal. This observation in one dataset
is investigated on a broader data basis in section~\ref{sec:SeizureDetection}
and a possible application of DyCA as seizure detection algorithm is discussed.

Considering the ODE~\eqref{eq:ShilnikovChaos} as a model of the epileptic seizure,
two eigenvalues are expected to be close to the value of $1$. Fig.~\ref{fig:WindowedDyCAEpilepsy}
shows this behavior within the seizure: the blue and the red line representing the
two largest eigenvalues are clearly closer to the value of 1 than the third eigenvalue.
By choosing an appropriate threshold, DyCA allows for an identification of projection
vectors leading to amplitudes which obey a linear set of ODEs.

Fig.~\ref{fig:ShilnikovChaosHomoclinic} shows on the right hand side the three-dimensional
trajectory of the amplitudes corresponding to the eigenvectors $u_1$ and $u_2$ and vector $v_1$.
The structure of the trajectory is clearly observable and the typical homoclinic orbit
of Shilnikov chaos is shown. DyCA represents a significant improvement compared to the
3D-trajectories in phase space spanned by the projection onto the first three PCA vectors
or the best ICA vectors.

\section{Detection of seizure events}
\label{sec:SeizureDetection}%

We now investigate the utilization of the DyCA-eigenvalues for the
detection of epileptic seizure events. For this a moving window frame
runs over the data and the DyCA eigenvalues are calculated on the
current window. The largest eigenvalues are compared against a
threshold. If the eigenvalues are larger than the threshold, it is
assumed that the current window contains an epileptic seizure.

To measure the quality of the classification method specificity (SPC),
false discovery rate (FDR) and miss rate/false negative rate (FNR)
were calculated. The specificity is the number of windows correctly
classified as not containing a seizure relative to the number of
windows not containing a seizure. The false discovery rate is the
ratio of windows falsely classified as containing a seizure over all
windows classified as containing a seizure. The miss rate is the
number of windows falsely classified as not containing a seizure in
relation to the number of all windows containing a seizure.

In our test setting DyCA detection was applied on six EEG data sets of
patients with absences, which are a special kind of epileptic
seizures. The mean length of the data sets is 411 seconds containing
absences of length ranging from 4 to 25 seconds. The size of the moving
window was taken as three seconds. As step size
for the movement of the window 10~\% of the window size were
taken. This results in windows having 90~\% overlap.
A window was labeled
as seizure if the whole window was contained in the pre-labeled
seizure. That is windows contained only partly in the seizure are
labeled as not being in the seizure. This results in a mean prevalence
of $2.32 \pm 2.18~\%$. No further pre-processing techniques, like
filtering, were applied to the data.

\begin{figure}
    \centering
    \includegraphics[width=0.45\textwidth]{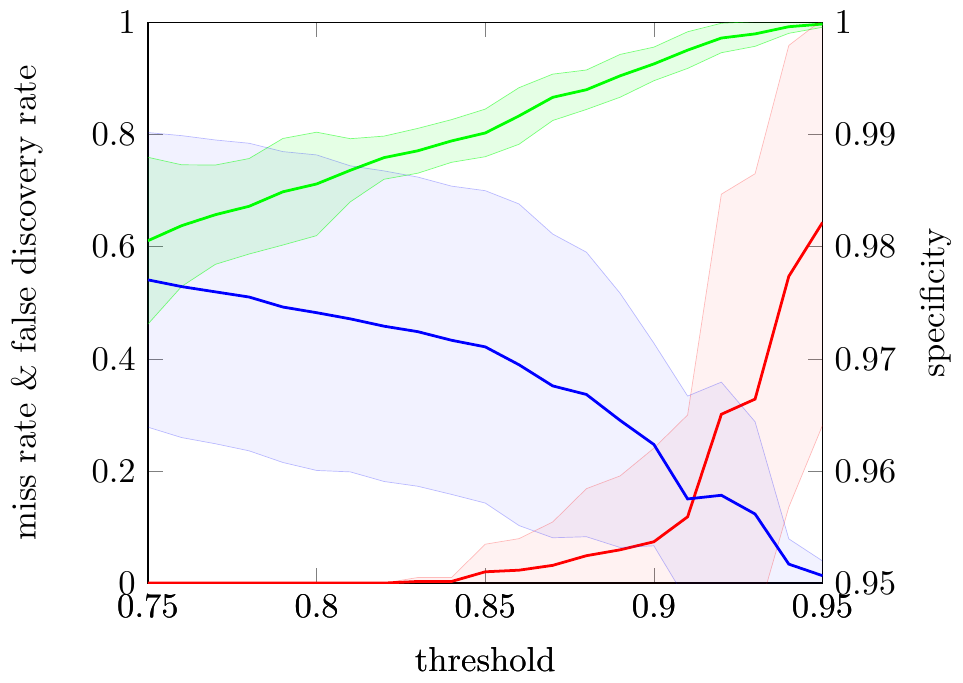}
    \caption{The specificity (green), false discovery rate (blue), and
      miss rate (red) plotted against the threshold for the largest
      eigenvalue. The shaded area shows the standard deviation with
      respect to different data sets.}
    \label{fig:1lambda}
\end{figure}

\begin{figure}
    \centering
    \includegraphics[width=0.45\textwidth]{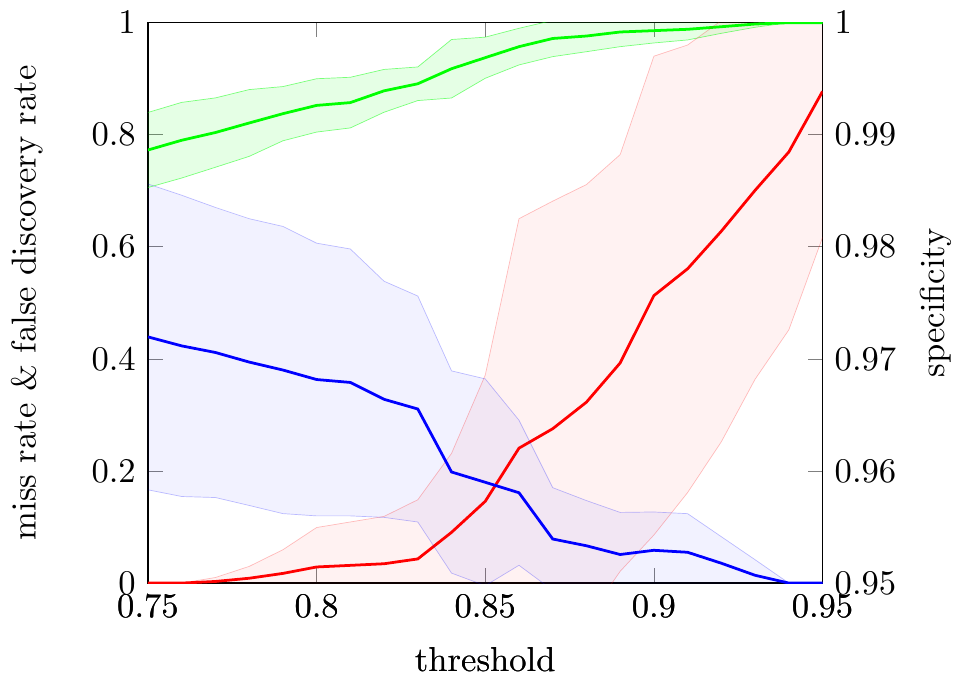}
    \caption{The specificity (green), false discovery rate (blue), and
      miss rate (red) plotted against the threshold for the two
      largest eigenvalues. The shaded area shows the standard
      deviation with respect to different data sets.}
    \label{fig:2lambda}
\end{figure}

On each window the DyCA eigenvalues were compared to a threshold. In
Fig.~\ref{fig:1lambda} the SPC, FDR, and FNR are shown for
classification if only the largest eigenvalue is compared against the
threshold. Analogously Fig.~\ref{fig:2lambda} shows SPC, FDR, and FNR
for classification comparing the two largest eigenvalues against
the threshold.


\section{Discussion}
\label{sec:Discussion}%

Considering the small prevalence the results in Fig.~\ref{fig:1lambda}
and Fig.~\ref{fig:2lambda} indicate that the two largest eigenvalues
should be considered, since then the specificity reaches 99~\%. This supports
from a data-driven point of view the theoretical considerations in
\cite{vanVeen.Liley:2006}. The choice of the threshold can be adapted
to the application in mind. If the detection of all seizures as soon as
possible is wanted and false alarms are acceptable the threshold should 
be lowered. If one is only interested in finding examples of appearing seizures
a higher threshold should be chosen. The high standard deviation and
relative high values of the false discovery rate is due to the labeling
process of the windows. Since requiring that the whole window is
contained in the seizure, windows containing only small parts outside
the seizure are labeled as no seizure even though the deterministic part
might be dominant. 

The detection algorithm can be adjusted to obtain a specificity of nearly
99.7~\% and still have a miss rate of 20~\%. 

\section{Conclusion}
\label{sec:Conclusion}%

In this paper we presented dynamical component analysis (DyCA) as an
alternative to PCA and ICA reducing the dimensionality of multi-variate
time series based on the assumption of an underlying dynamical system.
Comparing the 3D-trajectories in phase space we obtained by DyCA more
obvious structures than obtained by PCA and ICA (Fig.~\ref{fig:ShilnikovChaosHomoclinic}).

Applying DyCA to EEG data of epileptic seizures we implemented a novel
seizure detection algorithm and obtained good results (specificity of
nearly 99.7~\%) in comparison with other studies using more complex
tools for detection: In \cite{Acharya:2018} using a deep convolutional
neural network for detection, a specificity of only 90~\% was reached.
Further studies used wavelet transform or wavelet packet decomposition
combined with ICA~\cite{Acharya:2012ICA} or PCA~\cite{Acharya:2012PCA}
and different classifiers. They reach a specificity of 97~\% with ICA
and 99~\% with PCA as intermediate step.

Generally, it might be advantageous for some applications to use DyCA
instead of ICA or PCA as intermediate step for more refined classification
methods. This is subject of work in progress and will be presented in
upcoming papers.

%\vfill
%\pagebreak

\vfill\pagebreak

\bibliographystyle{IEEEbib}
\bibliography{Literatur}

\end{document}

%% file: Macros.tex
\renewcommand{\mathbb}[1]{\mathbbm{#1}}

\allowdisplaybreaks

\let\originalleft\left
\let\originalright\right
\renewcommand{\left}{\mathopen{}\mathclose\bgroup\originalleft}
\renewcommand{\right}{\aftergroup\egroup\originalright}

\newcommand{\argument}       {\,\cdot\,}

\DeclarePairedDelimiter{\timeavg}{\langle}{\rangle_t}

\DeclarePairedDelimiter{\norm}{\lVert}{\rVert}

%% file: ms.bbl
\begin{thebibliography}{1}

\bibitem{Ahmadi:2017}
A.~Ahmadi, V.~Shalchyan, and M.~R. Daliri,
\newblock ``A new method for epileptic seizure classification in {EEG} using
  adapted wavelet packets,''
\newblock {\em 2017 Electric Electronics, Computer Science, Biomedical
  Engineerings' Meeting (EBBT)}, pp. 1--4, 2017.

\bibitem{Acharya:2012PCA}
U.~R. Acharya, S.~V. Sree, A.~P.~C. Alvin, and J.~S. Suri,
\newblock ``Use of principal component analysis for automatic classification of
  epileptic {EEG} activities in wavelet framework,''
\newblock {\em Expert Systems with Applications}, vol. 39, no. 10, pp. 9072 --
  9078, 2012.

\bibitem{Acharya:2012ICA}
U.~R. Acharya, R.~Yanti, G.~Swapna, V.~S. Sree, R.~J. Martis, and J.~S. Suri,
\newblock ``Automated diagnosis of epileptic electroencephalogram using
  independent component analysis and discrete wavelet transform for different
  electroencephalogram durations,''
\newblock {\em Proceedings of the Institution of Mechanical Engineers, Part H:
  Journal of Engineering in Medicine}, vol. 227, no. 3, pp. 234--244, 2013.

\bibitem{Acharya:2018}
U.~R. Acharya, S.~L. Oh, Y.~Hagiwara, J.~H. Tan, and H.~Adeli,
\newblock ``Deep convolutional neural network for the automated detection and
  diagnosis of seizure using {EEG} signals,''
\newblock {\em Computers in Biology and Medicine}, vol. 100, pp. 270 -- 278,
  2018.

\bibitem{Pearson:1901}
K.~Pearson,
\newblock ``{On lines and planes of closest fit to a system of points in
  space},''
\newblock {\em The London, Edinburgh, and Dublin Philosophical Magazine and
  Journal of Science}, vol. 6,2, pp. 559--572, 1901.

\bibitem{Hyvaerinen.Oja:2000a}
A.~Hyvärinen and E.~Oja,
\newblock ``{Independent component analysis: algorithms and applications},''
\newblock {\em Neural Networks}, vol. 13, no. 4-5, pp. 411--430, 2000.

\bibitem{Seifert.Korn.Hartmann.Uhl:2018a}
B.~Seifert, K.~Korn, S.~Hartmann, and C.~Uhl,
\newblock ``{Dynamical Component Analysis (DyCA): dimensionality reduction for
  high-dimensional deterministic time-series},''
\newblock in {\em Proceedings of the IEEE 28th International Workshop on
  Machine Learning for Signal Processing (MLSP2018)}, 2018,
\newblock to appear.

\bibitem{vanVeen.Liley:2006}
L.~van Veen and D.T.J. Liley,
\newblock ``{Chaos via Shilnikov's Saddle-Node Bifurcation in a Theory of the
  Electroencephalogram},''
\newblock {\em Phys. Rev. Lett.}, vol. 97, pp. 208101, 2006.

\bibitem{Friedrich.Uhl:1996a}
R.~Friedrich and C.~Uhl,
\newblock ``Spatio-temporal analysis of human electroencephalograms: Petit-mal
  epilepsy,''
\newblock {\em Physica D}, vol. 98, no. 1, pp. 171--182, 1996.

\end{thebibliography}
